\documentclass[%
 reprint,
superscriptaddress,
 amsmath,amssymb,
 aps,
 prl,
]{revtex4-2}
\usepackage{aas_macros}
\usepackage{graphicx}
\usepackage{dcolumn}
\usepackage{bm}
\usepackage[normalem]{ulem}
\usepackage{hyperref}

\usepackage{tikz}
\usetikzlibrary{shapes.geometric, arrows, positioning}
\usetikzlibrary{shapes.geometric, arrows.meta, positioning, shapes}
\usepackage{xcolor}
\definecolor{mypurple}{rgb}{0.5, 0, 0.85}

\begin{document}

\preprint{APS/123-QED}

\title{Detecting Gravitational Wave Memory in the Next Galactic Core-Collapse Supernova}

\author{Colter J. Richardson}
\email{cricha80@vols.utk.edu}
\affiliation{Department of Physics and Astronomy, University of Tennessee, Knoxville, TN 37996, USA}
\author{Haakon Andresen} 
\affiliation{The Oskar Klein Centre, Department of Astronomy, AlbaNova, SE-106 91 Stockholm, Sweden}
\author{Anthony Mezzacappa}
\affiliation{Department of Physics and Astronomy, University of Tennessee, Knoxville, TN 37996, USA}
\author{Michele Zanolin}
\affiliation{Embry-Riddle Aeronautical University, 3700 Willow Creek Road, Prescott, Arizona 86301, USA}
\author{Michael G. Benjamin}
\affiliation{Department of Physics and Astronomy, University of Tennessee, Knoxville, TN 37996, USA}
\author{Pedro Marronetti}
\affiliation{Physics Division, National Science Foundation, Alexandria, Virginia 22314, USA}
\author{Eric J. Lentz}
\affiliation{Department of Physics and Astronomy, University of Tennessee, Knoxville, TN 37996, USA}
\affiliation{Physics Division, Oak Ridge National Laboratory, P.O. Box 2008, Oak Ridge, Tennessee 37831-6354, USA}
\author{Marek~J.~Szczepa\'nczyk}
\affiliation{Faculty of Physics, University of Warsaw, Ludwika Pasteura 5, 02-093 Warszawa, Poland}

\date{\today}

\begin{abstract}
We present an approach to detecting (linear) gravitational wave memory in a Galactic core-collapse supernova {\em using current interferometers}. Gravitational wave memory is an important prediction of general relativity that has yet to be confirmed. Our approach uses a combination of Linear Prediction Filtering and Matched-Filtering. We present the results of our approach on data from core-collapse supernova simulations that span a range of progenitor mass and metallicity. We are able to detect gravitational wave memory out to 10 kpc. We also present the False Alarm Probabilities assuming an On-Source Window compatible with the presence of a neutrino detection. 
\end{abstract}
\maketitle

\paragraph{Introduction}
The deaths of massive stars in core-collapse supernovae (CCSNe) are promising sources of gravitational waves (GWs). Stellar core collapse, core bounce at super-nuclear densities, fluid instabilities in the newly-formed proto-neutron star and in the cavity between the proto-neutron star surface and the supernova shock wave, believed to be vital to the explosive central engine, as well as explosion itself and anisotropic neutrino emission, are all expected to generate GWs \cite{KoKu17,AbPaRa22,MeZa24}. The fluid instabilities, as well as the turbulence they induce, are expected to excite GW emission at frequencies between 50\,Hz and a few kHz
\citep{
Mueller_1982,
MoScMu91,
MuJa97,
DiFoMu01,
KoYaSa03,
KoYaSa04,
MuRaBu04,
KoIwOh09,
MaJaMu09,
MuOtBu09,
Scheidegger_2010,
KoIwOh11,
MuJaWo12,
CeDeAl13,
MuJaMa13,
Ott_2013,
KuTaKo14,
YaMeMa15,
HaKuKo15,
HaKuNa16,
KuKoTa16,
KuKoHa17,
RiOtAb17,
Andresen_2017,
MoRaBu18,
OCCo18,
TaKo18,
HaKuKo18,
KaKuTa18,
Kuroda_2018,
Pan_2018,
AnMuJa19,
RaMoBu19,
VaBuRa19,
SrBaBr19,
Powell_2019,
MeMaLa20,
PoMu20,
ShKuKo20,
Warren_2020,
VaBu20,
AnGlJa21,
AnZhda21,
PaWaCo21,
ShKuKo21,
Pan_2021,
KuFiTa22,
MaAsTa22,
Nakamura_2022,
PoMu22,
BuGuFo23,
DrAnDi23,
JaMuHe23,
KuSh23,
MeMaLa23,
PaVaPa23,
VaBuWa23,
AfKuCa23,
BrBiOb23,
LiRiLu23,
RiZaAn23}.

Detection strategies for CCSN GWs until now relied on excess-energy methods because the stochastic nature of the signals impeded the use of matched filtering. However, it has been pointed out recently that matched filtering alongside multi-messenger observations can improve the detection efficiency of nearby events \cite{DrAnDi23}. Besides emissions above 50 Hz, a slowly evolving signal component, associated with the  GW (linear) memory, is expected below a few 10's of Hz 
\cite{
Epstein_1978,
Turner_1978,
MuJa97,
MuRaBu04,
KoIwOh09,
MuOtBu09,
YaMaMe10,
KoIwOh11,
TaKo11,
MuJaWo12,
MuJaMa13,
YaMeMa15,
MoRaBu18,
PoMu19,
RaMoBu19,
PoMu20,
JaPoMu22,
NaTaKo22,
Richardson22,
MeMaLa23,
PaVaPa23,
PoMuAg23,
VaBuWa23,
MuCaLu21,
MuLiLu22}.
The memory in a CCSN stems from asymmetric emission of neutrinos during the explosion and the non-spherical expansion of the supernova blast wave. Although this low-frequency component contributes minimally to the total energy emitted, its amplitude can be several times larger than that of the emission above 50 Hz. Strictly speaking, the memory only refers to a constant offset in the strain after the GW pulse has passed. However, in our discussion of the memory we include the secular ramp-up to the saturation value. In terms of detectability, the GW memory from CCSNe has been largely overlooked due to the limited sensitivity of current GW detectors below 10 Hz. Moreover, even if the peak of the frequency band of the memory (including the secular ramp-up) is below 10 Hz, there may be a detectable strain (or energy) present at and above 10 Hz.

In this Letter, we demonstrate that the slow and regular time evolution of the memory is uniquely suited to matched-filter techniques. We show how matched-filtering can be utilized to detect the GW memory from CCSNe {\em in current interferometers}. Observing the memory, or signs of it, would confirm an important prediction of general relativity that has yet to be confirmed. 

\paragraph{Models}
We study the memory from three state-of-the-art, three-dimensional core-collapse supernova simulations. The simulations were carried out with the \textsc{Chimera} \cite{CHIMERA} code, initiated from three non-rotating progenitors with zero-age main sequence masses of 9.6, 15, and 25 Solar masses, and zero and Solar metallicity \cite{MeMaLa23}. The models are labeled by a ``D'' (for \textsc{Chimera} D-series simulations) followed by the mass of the progenitor from which the simulation in the series was initiated. Rapid shock expansion sets in at $\sim$125\,ms, $\sim$250\,ms, and $\sim$500\,ms for D9.6, D25, and D15, respectively.

\begin{figure}
    \centering
    \includegraphics[width=0.5\textwidth]{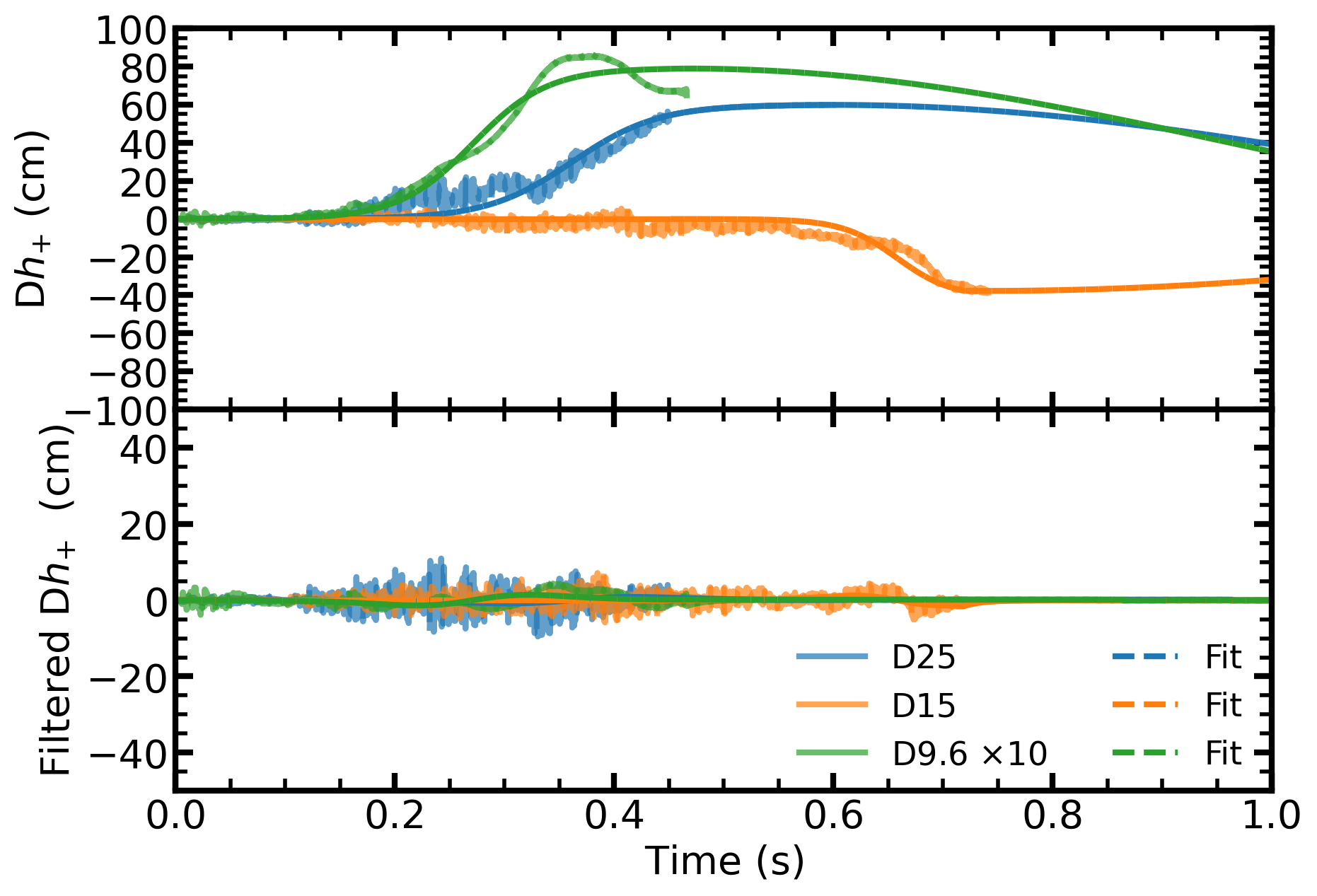}
    \caption{Gravitational wave signals from all three models. Solid lines show the signals from the simulations. Dashed lines show the fit of the waveforms to a logistic function with a tapering (see Eq.~\ref{eq:fit} and Table \ref{tab:logistic_parameters}). 
    Tapering reduces the high-frequency noise induced by the abrupt end of the simulations.
    We taper with a frequency of 1/10 Hz. The top panel displays unfiltered signals; the bottom panel shows the same signals with an 8 Hz high-pass Butterworth filter applied. 
    The D9.6 signal is scaled up by a factor of 10.}
    \label{fig:gwsignals}
\end{figure}

\paragraph{Gravitational Wave Signals}
The solid lines in the top panel of Fig.~\ref{fig:gwsignals} show the plus polarization mode of the combined (matter and neutrino) GW strains ($h_{+}$) from our three models: the blue, orange, and green curves represent D25, D15, and D9.6, respectively. Within a spherical coordinate system centered on the simulations, the models are observed at randomly-chosen directions: D9.6 at $\phi = -35^\circ$, $\theta = 90^\circ$; D15 at $\phi = 60^\circ$, $\theta = 70^\circ$; and D25 at $\phi = 35^\circ$, $\theta = 0^\circ$. The GW signals of all three models show the slow ramp-up to a non-zero strain value that is characteristic of the memory. (Note, the D9.6 model is representative of low-mass CCSNe, which typically have low ejecta asymmetry.)

\paragraph{Analytical Fit}

GW signals from CCSN explosions are stochastic. However, the ramp-up and memory phases exhibit a high degree of regularity that can be modeled using simple logistic functions, making them templatable.

To isolate the memory, we fit the signal to a tapered logistic function defined as follows:
\begin{align} \label{eq:fit}
    h^{\rm fit}(t) &= \frac{L}{1+e^{-k (t - t_{0})}} (1 - H(t-t_{s})) \nonumber \\
    &+\frac{L}{2} \big(1 + \cos{(2 \pi f_{t} (t - t_{s})})\big) H(t-t_{s}).
\end{align}
Here $t_{0}$ is the center of the rise time, $k$ is the inverse of the typical rise time, $L$ is the memory saturation value, $t_s$ is the time of saturation (or the end time of the simulation), and $H(t)$ is the Heaviside step function. The tapering is characterized by the tapering frequency, $f_t$, which we chose based on the noise characteristics of the LIGO-Virgo-Kagra (LVK) detectors. Current ground-based detectors have a sharp increase in their characteristic noise at approximately 10 Hz. As long as the tapering is longer than 0.1\,s, the small amount of energy added by the tapering is negligible in the sensitive band above 10 Hz used in the searches. In reality, the signals are expected to saturate at some non-zero value, but we taper the signals to avoid inducing high-frequency noise in our Fourier analysis. The dashed curves in the top panel of Fig.~\ref{fig:gwsignals} show the fits. See Table~\ref{eq:fit} for the fit parameters. For the D9.6 and D15 models, we start the tapering right as the simulations end. On the other hand, we extrapolate the D25 signal until it saturates (details regarding extrapolating the signals can be found in \cite{Richardson22}). We extrapolate the signal from D25 because applying the tapering directly after the end of the simulation led to a discontinuity in the signal's derivative. The extrapolation is conservative and is not instrumental to detecting the actual signal. 

\begin{table}
    \centering
    \caption{The parameters used for the fit and tapering of the GW signals from the simulations (see Eq.\ref{eq:fit}). Each row corresponds to a particular model.}
    \begin{tabular}{cccccc}
        \hline\hline
        Model & $t_{0}\ [s]$ & $L\ [cm]$ & $k\ [Hz]$ & $t_s\ [s]$ & $f_t\ [Hz]$\\
        \hline
        D9.6  & 0.27  & 7.92 & 29.33 & 0.300 & 0.1 \\
        D15   & 0.68 & -40.96 & 40.01 & 0.7414 & 0.1 \\
        D25   & 0.36 & 60.0 & 25.6 & 0.472 & 0.1 \\
        \hline\hline
    \end{tabular}
    \label{tab:logistic_parameters}
\end{table}

\paragraph{Matched Filtering}

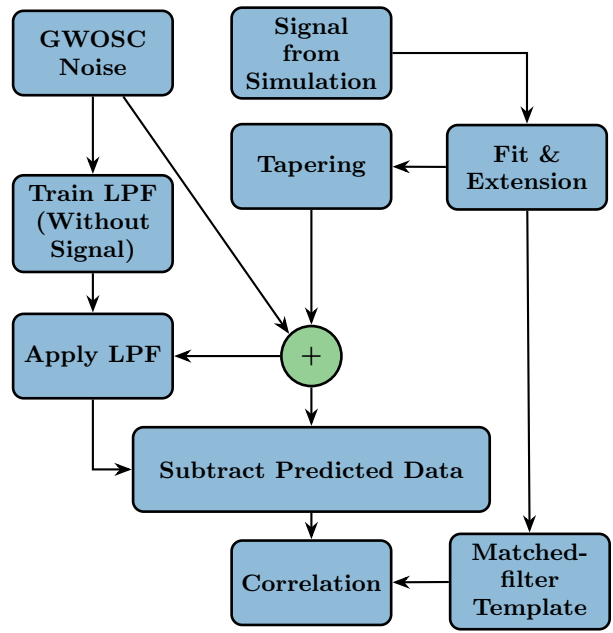
\begin{figure}
\begin{tikzpicture}[font=\small,>=Stealth,line width=1pt]
\definecolor{customblue}{HTML}{1f77b4}
\definecolor{customgreen}{HTML}{2ca02c}
    \bfseries
    \tikzstyle{block} = [rectangle, draw, fill=customblue!50, text width=5em, text centered, rounded corners, minimum height=3em]+    \tikzstyle{wblock} = [rectangle, draw, fill=customblue!50, text width=12em, text centered, rounded corners, minimum height=3em]
    \tikzstyle{line} = [->, thick]
    \tikzstyle{smcirc} = [circle, draw, fill=customgreen!50, text width=1em, text centered, rounded corners]    
    \node [block] (gwosc) {GWOSC Noise};
    \node [block, right of = gwosc, xshift=0.75in] (sim) {Signal from \\Simulation};
    \node [block, below right of = sim, yshift=-0.32in,xshift=0.85in] (ext) {Fit \& \\Extension};
    \node [block, below of = ext, yshift=-0.6in] (tap) {Tapering};
    \node [block, below of=gwosc, yshift = -0.5in](lpf) {Train LPF (on Signal-less Data)};
    \node [smcirc, below of=sim, yshift=-01.2in] (plus) {+};
    \node [block, below of=lpf, yshift = -0.3in](alpf) {Apply LPF};
    \node [wblock, below of=plus, yshift=-0.2in] (rlpf) {
    Subtract Predicted Data};
    \node [block, below of=rlpf, yshift=-0.2in] (cor) {Correlation};
    \node [block, below of=ext, right of=cor, xshift=0.75in, yshift=0.4in] (fil2) {Matched-filter Template};
    \draw [line] (gwosc) -- (lpf);
    \draw [line] (gwosc) -- (plus.north west);
    \draw [line] (sim)   -|  (ext);

    \draw [line] (tap)   --  (plus);

    \draw [line] (ext)   --  (tap);

    \draw [line] (tap)   --  (fil2);
    \draw [line] (fil2)   --  (cor);
    \draw [line] (rlpf)   --  (cor);
    \draw [line] (plus)   --  (rlpf);
    \draw [line] (lpf)   --  (alpf);
    \draw [line] (plus)   --  (alpf);
    \draw [line] (alpf)   |-  (rlpf);
\end{tikzpicture}
    \caption{Flow chart outlining the procedure presented in this Letter. The end node labeled ``Correlation'' corresponds to the final result of our analysis and is what we show in Fig. \ref{fig:injected1}.}
    \label{fig:flowchart}
\end{figure}

The procedure implemented in this Letter, starting from the GWOSC noise data and the waveforms predicted by our simulations, is outlined in Fig.~\ref{fig:flowchart}. 

We inject the tapered signals into a sample of LVK data obtained from the Gravitational Wave Open Science Center (GWOSC) \citep{LVKgwosc}---specifically, a 4096\,s segment of data from the O3b run of the Livingston and Hanford detectors, with an initial GPS time of 1262178304. 
This 4096s window was chosen because it represents a single data frame from GWOSC.
Due to the nature of the publicly available data from GWOSC, which has a high-pass filter already applied to it, after injection we apply to the strain a high-pass Butterworth filter with a cut-off of 8\,Hz. 
This high-pass Butterworth filter was used to mimic the filter applied by the LVK Scientific Collaboration to remove everything below 10 Hz. If the filter were not applied, the signal-to-noise ratio would have been boosted artificially in this frequency band. The LIGO noise above 10 Hz was not altered for a filter cutoff of 8 Hz.

The second panel of Fig.~\ref{fig:gwsignals} shows the signals and the fits after the filter has been applied. For all of the models, the secular ramp-up is reduced, but not erased.

After injecting the signal, we train a Linear Prediction Filter (LPF) \cite{Jackson1996} with 16384 trained parameters on a 2048\,s segment of the data that does not contain the signal. We tested training the LPF using both the \textit{lpc} function in MATLAB \cite{MATLAB} and the Python module \textit{librosa}  \cite{mcfee15}. {Both implementations showed similar results, but we used MATLAB for this work.}
The LPF was trained using the half of the data not containing the signal, ensuring that the training data was both sufficiently long and signal free. The training converged for training segments
shorter than the 2048 seconds used here.
We then subtract the portion of the signal predicted by the LPF and define
\begin{align} \label{eq:filterd_strain}
    \hat{S} = S - S_{\rm LPF},
\end{align}
where $S$ is the strain from the detector, including the injected signal, and $S_{\rm LPF}$ is the output of the LPF. 
\begin{figure}
    \centering
    \includegraphics[width=0.5\textwidth]{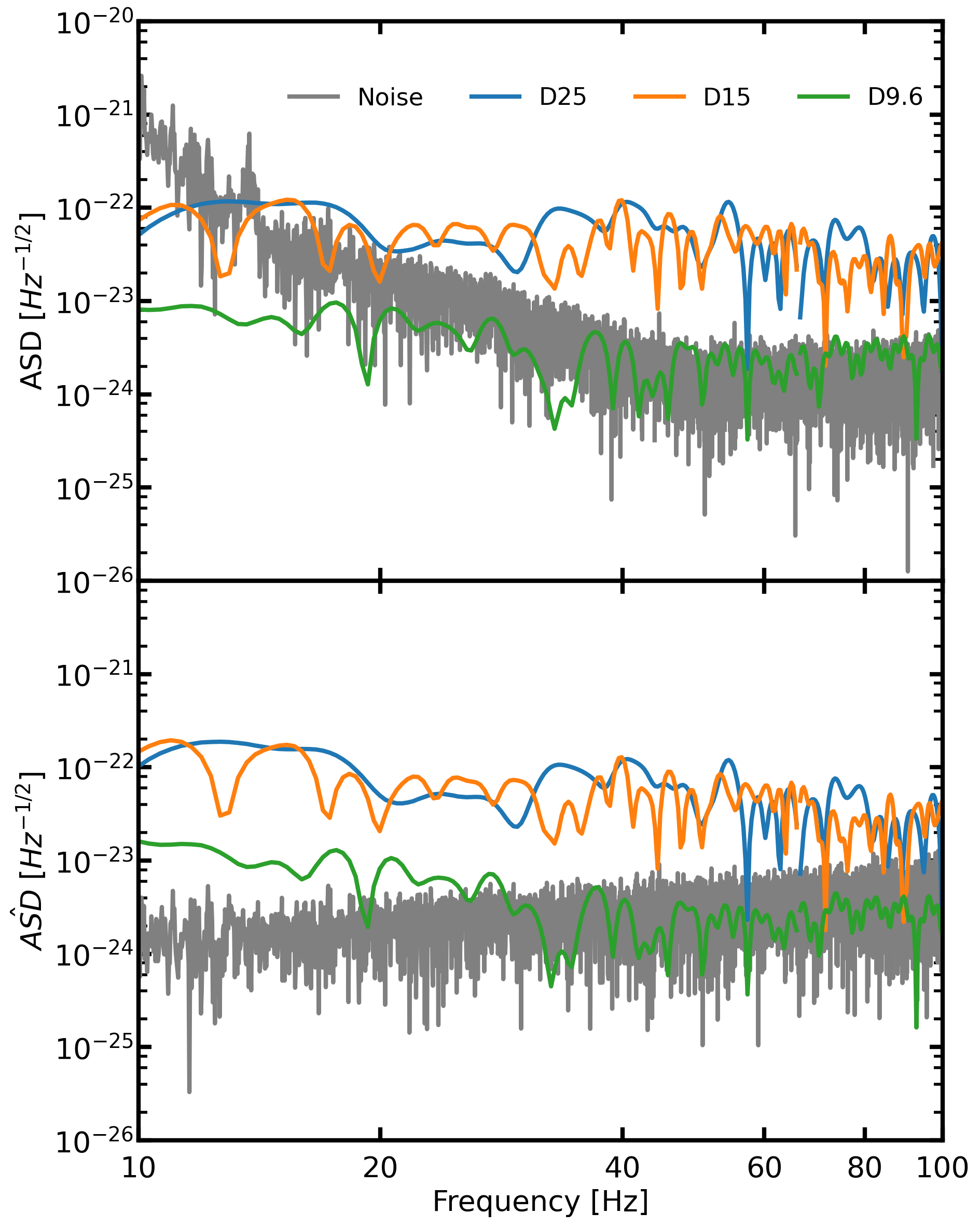}
    \caption{Top: The Amplitude Spectral Density of (1) the noise data from the Livingston detector (gray) and (2) the three signals (colored lines). The inset plot shows the convergence of the LPF as a function of training time. Bottom: The Amplitude Spectral Density of (1) the detector noise with the part predicted by the LPF subtracted (gray) and (2) the whitened signals (colored lines). Signals were scaled to a source distance of 1 kpc.}
    \label{fig:psd}
\end{figure}
In Fig.~\ref{fig:psd} we show the amplitude spectral density (ASD) of the noise and the signals. The top (bottom) panel shows the data before (after) applying the LPF. The data in the bottom panel are defined in Eq.~\ref{eq:filterd_strain}, and, for the purpose of the plots only, we assume a source distance of 1\,kpc. The filtered data do not represent the actual detector strain, but by predicting, and then removing the predicted portion of the noise, we are able to better locate the memory component. The LPF below 200 Hz improves detectability metrics (like the SNR or cross-correlations with our templates) for the memory by several orders of magnitude. It does affect the signal as well, but to a far lesser degree. Therefore, $\hat{S}$ is a better starting point for our matched filtering than the actual strain data.
The change in amplitude after the application of the LPF is a concern only if one is trying to determine the amplitude of the signal. The LPF is designed to suppress the noise. Its impact on the signal can increase, decrease, or leave constant the signal amplitude, as seen in Fig.~\ref{fig:psd}. The L2-Norm of the noise with the contribution predicted from the LPF shows convergence at training times larger than 1000 seconds.

We then calculate the discrete correlation---i.e., the {\em match}---between $\hat{S}$ and $h^{\rm fit}$, which is defined as follows:
\begin{align} \label{eq:match}
    \langle\hat{S}, h^{\rm fit}\rangle(t_n) = \sum_{m} \hat{S}(t_{m}) h^{\rm fit}(t_{m-n}).
\end{align}
Here we sum over all $t_m$ in our data, and $t_n$ refers to the $n$-th sample time. We simulate a two-detector network by evaluating
\begin{align} \label{eq:match_network}
    \langle\hat{S}, h^{\rm fit}\rangle^{\rm N} = \langle\hat{S}, h^{\rm fit}\rangle^{\rm H} \cdot \langle\hat{S}, h^{\rm fit}\rangle^{\rm L},
\end{align}
where $\langle\hat{S}, h^{\rm fit}\rangle^{\rm H}$ and $\langle\hat{S}, h^{\rm fit}\rangle^{\rm L}$ represent $\langle\hat{S}, h^{\rm fit}\rangle$ calculated with data from the LIGO Hanford and LIGO Livingston detectors, respectively. A detector network reduces the number of false alarms by enabling coincident analysis. We leave a more complete exploration of the impact of detector networks to future work.

We split the 4096 s window into two-second segments and define the FAP as the ratio between the number of segments with triggers above some threshold and the total number of segments. The motivation for using
two-second windows is twofold. First,
the simulations investigated here, and those of many others 
\citep{
Mueller_1982,
MoScMu91,
MuJa97,
DiFoMu01,
KoYaSa03,
KoYaSa04,
MuRaBu04,
KoIwOh09,
MaJaMu09,
MuOtBu09,
Scheidegger_2010,
KoIwOh11,
MuJaWo12,
CeDeAl13,
MuJaMa13,
Ott_2013,
KuTaKo14,
YaMeMa15,
HaKuKo15,
HaKuNa16,
KuKoTa16,
KuKoHa17,
RiOtAb17,
Andresen_2017,
MoRaBu18,
OCCo18,
TaKo18,
HaKuKo18,
KaKuTa18,
Kuroda_2018,
Pan_2018,
AnMuJa19,
RaMoBu19,
VaBuRa19,
SrBaBr19,
Powell_2019,
MeMaLa20,
PoMu20,
ShKuKo20,
Warren_2020,
VaBu20,
AnGlJa21,
AnZhda21,
PaWaCo21,
ShKuKo21,
Pan_2021,
KuFiTa22,
MaAsTa22,
Nakamura_2022,
PoMu22,
BuGuFo23,
DrAnDi23,
JaMuHe23,
KuSh23,
MeMaLa23,
PaVaPa23,
VaBuWa23,
AfKuCa23,
BrBiOb23,
LiRiLu23,
RiZaAn23},
predict GW-memory ramp-up to occur in less than two seconds. While few predictions suggest longer ramp-up times, any signal developing over a period longer than two seconds would be buried in the noise of the detector, and also filtered out of the data by the high-pass filter. A two seconds window is, therefore, an appropriate choice for the current interferometric data and most memory models.
Second, for a Galactic CCSN, timing information based on a detected neutrino event would enable us to significantly narrow our search window to 
a few seconds.

We consider as a single GW candidate all the transient energy in a 2 seconds
window.  All searches for GW bursts have "clustering" procedures 
merging temporally close transients (cWB for example).
Internal tuning parameters in GW searches are expected 
to be reasonable, but ultimately their value lies in the search performance.

When performing matched filtering, the sharp edges of noise segments at the beginning and the end of a data stream can lead to large and nonphysical correlations (edge effects). Therefore, we apply a window to the noise data before our matched search. We used a Tukey window of the same length as our noise, with a shape parameter $\alpha = 0.2$. 

Lastly, we note that the signals are injected at a randomly-chosen time and that choosing a different injection time does not  change the general conclusions of this work. However, for the noise data we use, there are three clusters where the amplitude of the correlation, which is at its base an inner product, is large, at roughly 750\,s, 894\,s, and 1350\,s. Injecting the signals near the noise clusters slightly decreases the efficiency of our approach. For a source distance of 10 kpc, the False Alarm Probability (FAP) increases by a few percent, but signals can still be clearly identified.
We found that the noise clusters were
glitch-like features, not yet ruled out by 
the LIGO data characterization. It might
be possible to veto these noise events, but we leave this for future work.

\begin{figure}
    \centering
    \includegraphics[width=0.5\textwidth]{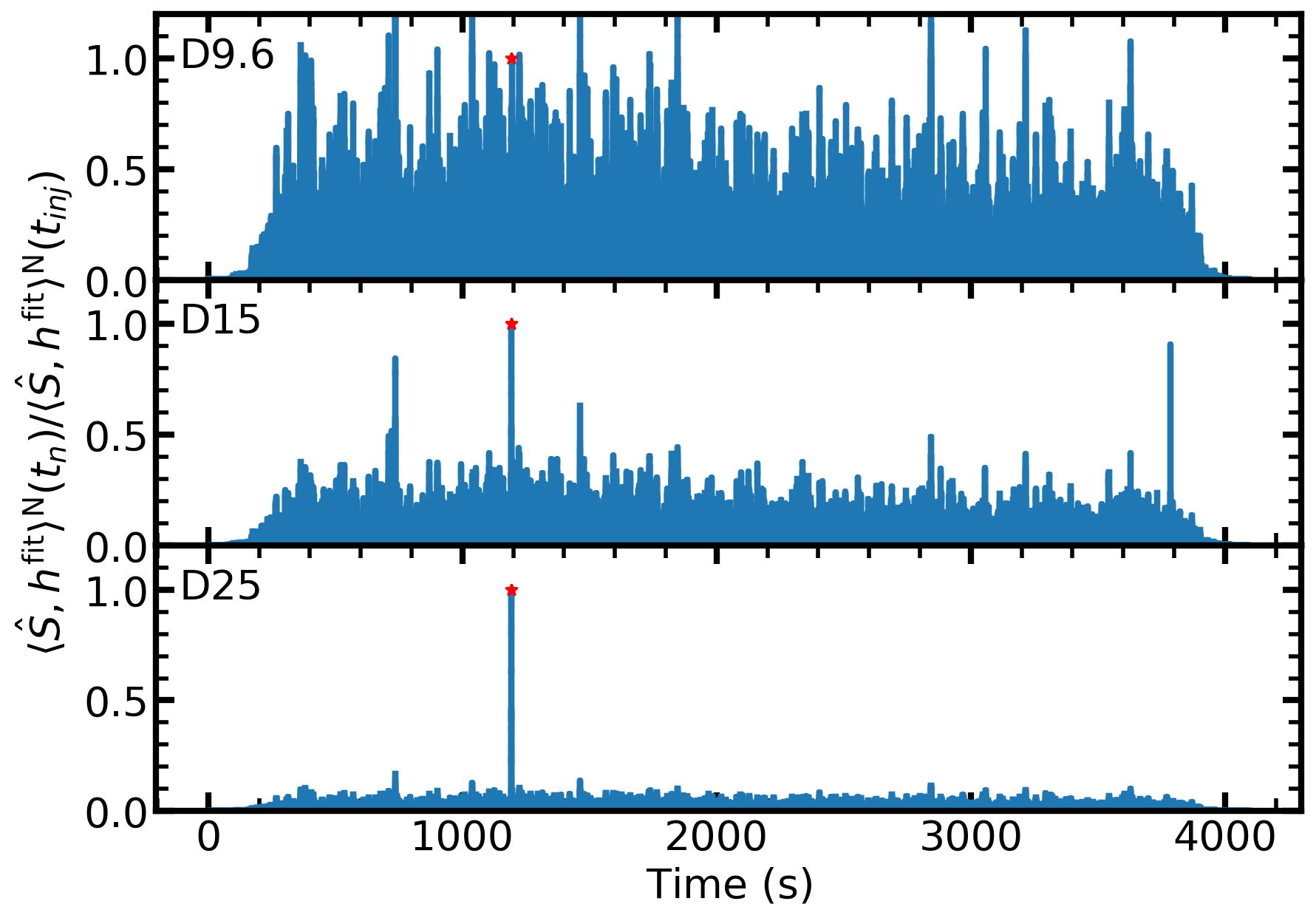}
    \caption{The two-detector correlation (Eq. ~\ref{eq:match_network}) between the filtered templates 
    (bottom panel of Fig~\ref{fig:gwsignals}) and the whitened detector data. 
    The top, middle, and bottom panels correspond to signals from the D9.6, D15, and D25 models, respectively, all at a distance of 10 kpc. 
    The star indicates the location of injection.} 
    \label{fig:injected1}
\end{figure}

\paragraph{Results}
The matched-filter results for all three models, with a source distance of 10 kpc and using a two-detector network at O3b sensitivity (for the expected design sensitivity of O5, we expect at least a 50\% increase), are presented in Fig.~\ref{fig:injected1}. The top, middle, and bottom panels correspond to models D9.6, D15, and D25, respectively. The signals were injected at one randomly-chosen time, indicated by gray dashed lines. The signals from D15 and D25 are identifiable, but the weak signal of D9.6 is not visible in the detector noise. In addition to the signal, our matched-filter approach picks out several other noise events (for example, around 750 and 2656 s, see Fig.~\ref{fig:injected1}). However, the correlations between the filtered template and noise events are smaller than the correlation between the template and the actual signal (except for D9.6 and a particularly loud noise event for D15). To calculate a FAP, we select a match threshold and only consider events with a correlation, $\langle\hat{S}, h^{\rm fit}\rangle^{\rm N}(t_{n}) / \langle\hat{S}, h^{\rm fit}\rangle^{\rm N}(t_{inj})$, higher than the chosen threshold as potential detection candidates. The threshold is chosen to achieve a desired FAP (for brevity, we leave for future studies the discussion of varying the template parameters while performing the matched filtering, even if our tests indicate that this will have a small impact on the FAP curves).

For the D15 signal injected at 10 kpc, the correlation has two distinct peaks, one at the injection site and one at the time of a glitch. This glitch is not present in either the D9.6 or the D25 case, indicating that this noise event only correlates with the D15 template. 
Using a threshold of 0.8, we see that the signal is one of two triggers, resulting in a FAP of 0.097\% (2/2048). 
Lowering the threshold to 0.5 results in six triggers, leading to a FAP of 0.293\%. For the D25 signal, applying the same 0.8 and 0.5 thresholds, we detect no false triggers at the higher threshold and one false trigger at the lower threshold, achieving FAPs of $\leq$ 1/2048 (0.05\%) and 0.097\%, respectively.
The FAPs reported here correspond to detections with a significance level above 3 sigma. At a distance of 10 kpc, the SNR of the D9.6, D15, and D25 models are approximately 1.2, approximately 21.0, and approximately 21.9, respectively.

GWOSC provides time intervals that should be discarded. In our analysis, we remove these intervals. However, they were not determined in the context of a memory search. Our results suggest that a specialized detector characterization for a memory search is warranted. 

The impact on the FAP of using a template bank for the memory (as opposed to the three templates used here) is expected to be small because, when a different ramp-up time is used, the matched-filter peaks will be of smaller amplitude and the new template would not be chosen as the representative template. (For example, the results from our D15 and D25 models shown in Fig.~\ref{fig:injected1} are consistent even though the ramp-up templates differ.)

\begin{figure}
    \centering
    \includegraphics[width=0.5\textwidth]{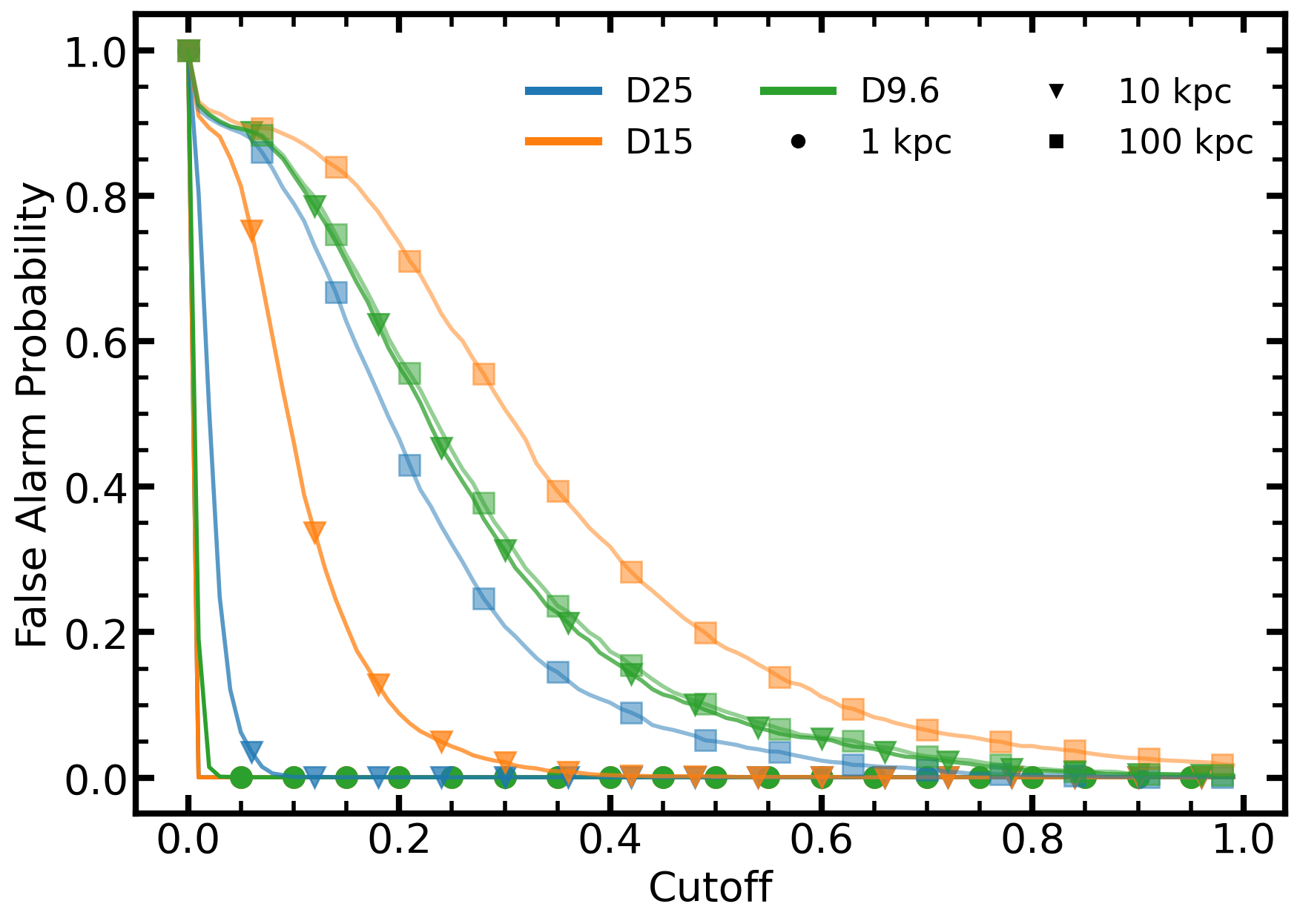}
    \caption{The False Alarm Probability as a function of the correlation threshold used to accept a trigger. Different line markers indicate different source distances, and the colors correspond to different models.}
    \label{fig:FAP}
\end{figure}

Given a coincident neutrino detection (or a search with a two-second temporal window), 
Fig.~\ref{fig:FAP} shows how the FAP depends on the chosen match threshold and the distance to the source. The blue, orange, and green curves correspond to models D25, D15, and D9.6, respectively. Different markers indicate different distances: $1$\,kpc (dots), $10$\,kpc (triangles), and $100$\,kpc (squares). At a distance of $1$\,kpc, the memory in the D15 and D25 models is detectable with a FAP less than 0.05\% for any match threshold, while for the D9.6 model, a large match threshold is required in order to obtain similar results.
(N.B. Our results at $O(1)$ kpc reflect the results we can expect at $O(10)$ kpc with next-generation detectors.) At $10$\,kpc, both D15 and D25 have a FAP of less than $5\%$  for a match threshold of 0.3. For D9.6, the FAP curve does not change relative to the 1 kpc case. This occurs when, at the injection distance, the signal becomes dominated by the noise, rendering the quantity plotted in Fig. \ref{fig:injected1} independent of the distance. This would eventually happen for the D15 and D25 models, as well, at some distance above 100 kpc.
At 100 kpc, a FAP for models D15 and D25 below 5-10\% is possible, but only at match thresholds approaching 1.0. At a threshold of 1.0, the FAP for both models is greater than 0.05\%, which means, at this distance, there are always triggers stronger than the injected signal. Note, the FAP curves for the D9.6 model and at large distances for models D15 and D25 (e.g., at 100 kpc) will vary with injection times, and will require potentially larger match thresholds at those distances.

The FAP presented in this Letter for a threshold of 0.8 shows that the D25 model can be identified at 10 kpc with a FAP $<$ 0.05\%, which is two orders of magnitude smaller than the 5\% threshold of interest currently used for gamma-ray-burst searches \cite{PhysRevLett.113.011102}. This, coupled with the noise floor reductions expected in next-generation detectors \cite{P2100003,2011CQGra..28i4013H}, shows the potential for detectability at Mpc distances with the same FAP thresholds. We plan to explore this direction in future publications. In this case a we would not have a neutrino detection and only a search mode triggered by an EM observation would be available. 

For our D25 model, assuming an optimally-orientated detector, a source distance of 10 kpc, without any data whitening, and using the LVK-O3 noise curve between 10 and 20 Hz, we obtain an SNR of 0.075. With the Einstein Telescope (ET) and Cosmic Explorer (CE) projected noise curves integrated from 5 Hz to 20 Hz and 1 Hz to 20 Hz, respectively, we obtain SNRs of 66.324 and 80.545, respectively, which indicate an improvement by at least two orders of magnitude.
It is not possible, with the 4096 s illustrative data set we use in this Letter, to probe smaller FAPs, for all distances, but we plan to do so in future investigations.

\paragraph{Conclusions}
In this Letter, we have shown that, given the secular ramp-up of the linear GW memory in a CCSN and given the use of a Linear Prediction Filter, a matched-template search (similar to the current detection strategy for binary mergers) can be performed to detect the memory {\em using current interferometers} and for the first time confirm an important prediction of general relativity. This is our primary finding.

With a focus on detecting CCSNe memory, we have shown that for current detectors our approach would be effective out to a distance of 10 kpc. Of course, at these distances a multimessenger detection is expected. The detection range afforded by our approach could also be considered in the context of next-generation interferometers. For the ET and CE, the combination of their reduced noise floor across all frequencies, which is projected to be approximately one order of magnitude in magnitude across the sensitivity band, and the reduction of the low-frequency wall from 10 Hz to below 20 Hz with improved controls noise, may enable the detection range of CCSNe---specifically, through the detection of GW memory---out to Mpc distance scales, necessarily without a concurrent neutrino detection, if a sufficiently low FAP could be achieved with OSWs of the order of days. In this case, the results from the matched filter alone, Fig. \ref{fig:injected1}, are to be expected.
A systematic exploration of this direction, including customized detector characterization, is left for a future publication. 

While this does not impact our main conclusions, in future publications we will discuss the variability of the results using different saturation levels for the memory (as a detection and parameter estimation template search). Recent findings suggest that we can have saturation values up to 60 times larger than our signals \cite{VaBuWa23}, and even larger for GRBs \cite{2023MNRAS.518.5242U}, potentially extending the detection range to several Mpc in current detectors. Additionally, it has been found that asymmetric neutrino emission can lead to large neutron star kicks, with potential amplification through neutrino flavor conversion \cite{NaSuYa19, NaSu24}. Such large, neutrino-induced kicks would imply a substantial GW memory, detectable at far greater distances than our current signals suggest.

\begin{figure}
    \centering
    \includegraphics[width=0.48\textwidth]{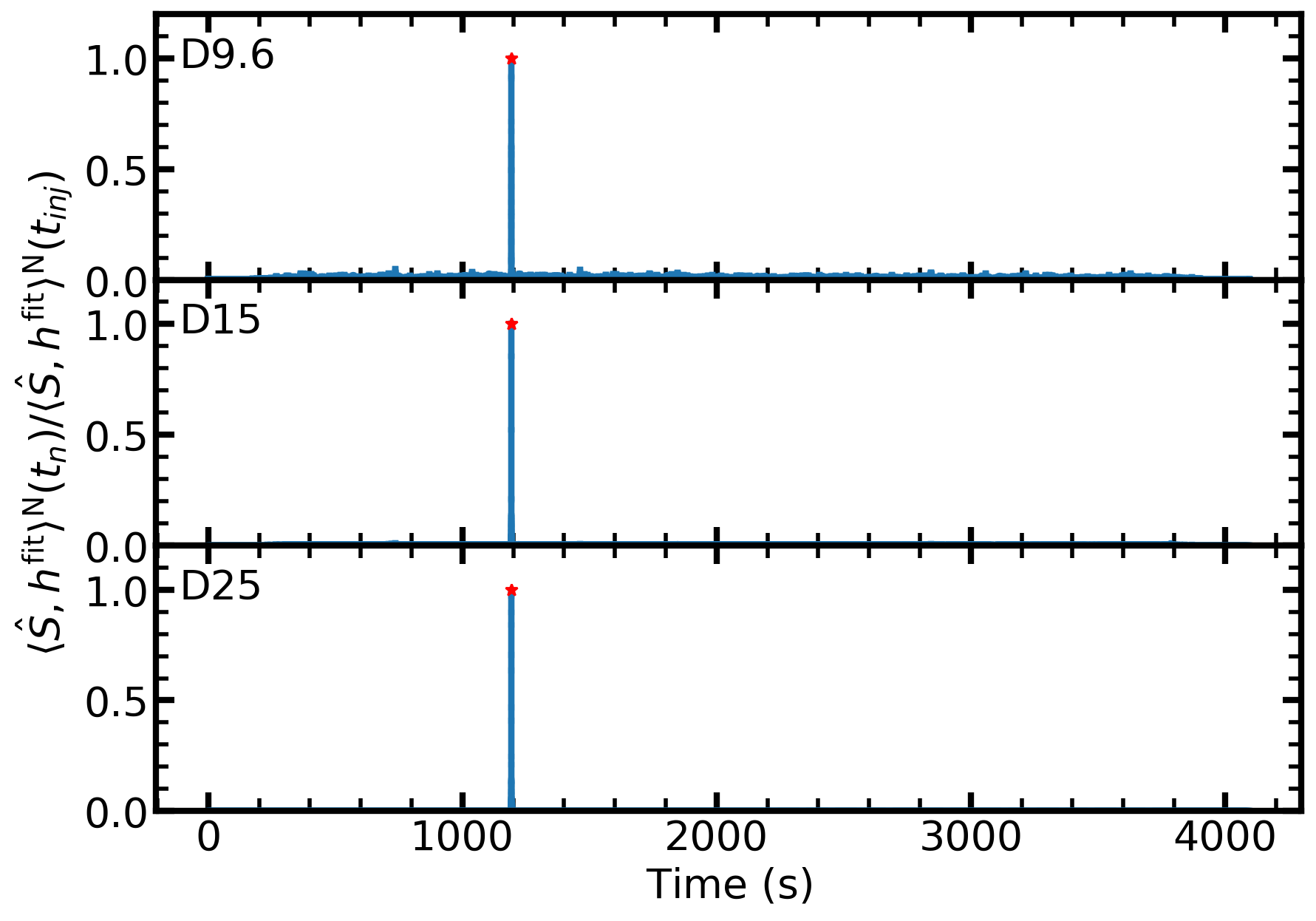}
    \caption{The two-detector correlation between the filtered templates (bottom panel of Fig. \ref{fig:gwsignals}) and the whitened detector data. The top, middle, and bottom panels correspond to signals from the D9.6, D15, and D25 models, respectively, all at a distance of 1 kpc. The star indicates the location of injection.}
    \label{fig:cor1}
\end{figure}

\paragraph{Acknowledgements}

H.A. is supported by the Swedish Research Council (Project No. 2020-00452).
A.M. acknowledges support from the National Science Foundation's Gravitational Physics Theory Program through grants PHY-1806692 and PHY-2110177. 
M.Z. is supported by the National Science Foundation Gravitational Physics Experimental and Data Analysis Program through awards PHY-2110555 and PHY-2405227. 
P.M. is supported by the National Science Foundation through its employee IR/D program. The opinions and conclusions expressed herein are those of the authors and do not represent the National Science Foundation.

This research has made use of data or software obtained from the Gravitational Wave Open Science Center (\href{https://gwosc.org}{gwosc.org}), a service of the LIGO Scientific Collaboration, the Virgo Collaboration, and KAGRA. This material is based upon work supported by NSF's LIGO Laboratory, which is a major facility fully funded by the National Science Foundation, as well as the Science and Technology Facilities Council (STFC) of the United Kingdom, the Max-Planck-Society (MPS), and the State of Niedersachsen/Germany for support of the construction of Advanced LIGO and construction and operation of the GEO600 detector. Additional support for Advanced LIGO was provided by the Australian Research Council. Virgo is funded, through the European Gravitational Observatory (EGO), by the French Centre National de Recherche Scientifique (CNRS), the Italian Istituto Nazionale di Fisica Nucleare (INFN) and the Dutch Nikhef, with contributions by institutions from Belgium, Germany, Greece, Hungary, Ireland, Japan, Monaco, Poland, Portugal, and Spain. KAGRA is supported by the Ministry of Education, Culture, Sports, Science, and Technology (MEXT), the Japan Society for the Promotion of Science (JSPS) in Japan, the National Research Foundation (NRF) and Ministry of Science and ICT (MSIT) in Korea, and the Academia Sinica (AS) and National Science and Technology Council (NSTC) in Taiwan.

 The authors would like to acknowledge \citet{2026PhRvD.113h3034L} for pointing out the discrepancy between our computed neutrino-induced waveforms and theirs, which prompted us to investigate the discrepancy and which led to the discovery of our error and corrections herein.

\bibliography{refs,3DGW,add_journals,additionalrefs,apj_journals,clean,paper,pr_add_journals}

\end{document}